\documentclass[journal,letterpaper]{IEEEtran}

\usepackage{graphicx,cite,epsfig,amssymb,amsmath,amsfonts,multicol,subfigure,mathtools,bm,mathrsfs,setspace}
\usepackage{multirow}
\usepackage{flushend}

\newcommand{\tabincell}[2]{\begin{tabular}{@{}#1@{}}#2\end{tabular}}

\begin{document}
\title{\huge A Method Beyond Channel Capacity in the Low SNR Regime:\\Theoretical Proof and Numerical Confirmation}
\author{
	Bingli~Jiao,
	Yuli~Yang
	and~Mingxi~Yin
	\thanks{B. Jiao ({\em corresponding author}) and M. Yin are with the Department of Electronics and Peking University-Princeton University Joint Laboratory of Advanced Communications Research, Peking University, Beijing 100871, China (email: jiaobl@pku.edu.cn, yinmx@pku.edu.cn).}
	\thanks{Y. Yang is with the Department of Electronic and Electrical Engineering, University of Chester, Chester CH2 4NU, U.K. (e-mail: y.yang@chester.ac.uk).}
}

\maketitle

\begin{abstract}

A method is proposed, referred to as the cocktail BPSK, to use two independent BPSK streams layered in a parallel transmission through the additive white Gaussian channel. The critical point of the method is to translate the interference between the two layered symbols into a problem of one symbol with two possible amplitudes, whereby we find that the achievable data rate (ADR) of the first layer equals to the channel capacity at zero signal-to-noise-ratio limit. By adding the contribution of the second layer, we find that the total ADR of the two layers can be larger than the channel capacity. The theoretical proof is done based on the mutual information and the numerical results are added for the confirmation.

\end{abstract}

\begin{IEEEkeywords}
Achievable data rate, channel capacity, cocktail BPSK, mutual information. 
\end{IEEEkeywords}

\IEEEpeerreviewmaketitle

\section{Introduction}

The aim of this correspondence is to present, from theoretical prospective, a method that can communicate higher realizable data rate beyond the channel capacity in the low signal-to-noise-ratio (SNR) regime. We referee the proposed method to as the cocktail BPSK because the two independent BPSK symbol-streams are adjusted in their amplitudes with the parallel transmission.  
 
A mathematical incentive can be found from 
\begin{eqnarray}\label{eq1}
\log_2(1+\rho) \leq \log_2(1+\rho_1) + \log_2(1+\rho_2),
\end{eqnarray}
with $\rho = \rho_1 + \rho_2$, due to the down-concave property of logarithm function [1]. Along with this thought [2], we work again by exploring the Euclidean geometrics to the signals' layering and splitting at the channel input and output respectively. 

Let us start from the mutual information 
\begin{eqnarray}\label{eqR}
\mathbb{R} = {I}(X;Y),
\end{eqnarray}
where $\mathbb{R}$ is the achievable data rate (ADR), $ I(X;Y)$ is the mutual information with the memoryless additive white Gaussian noise (AWGN) channel model
\begin{eqnarray}\label{eqy}
y = x + n,
\end{eqnarray}
where $y$ is the channel output, $x$ is the channel input, and $n$ is the AWGN component from a normally distributed ensemble of power $\sigma_N^2$, denoted by $n \sim \mathcal{N}(0,\sigma_N^2)$.

As has been well understood, maximizing \eqref{eqR} with respect to the input distribution yields the capacity
\begin{eqnarray}\label{eqC}
C = \mathop {\max }\limits_{p(x)} \{ I(X;Y)\} = \log_2(1 + \rho ) ,
\end{eqnarray}
where $C$ is the capacity measured in bits per second per Hz and $\rho$ is the SNR. 

The explicit expression of our derivations requires the theoretical perfection of capacity-achieving code with the upper bounds of \eqref{eqR} and \eqref{eqC} to perform the error free transmissions of the source codes without exhibiting any specific channel codes. Based on this, we assume that the transmitted information symbols, e.g., $x$ in \eqref{eqy}, can be correctly recovered at the receiver. 

The excellency of our theoretic result owns to the parallel transmission of two independent BPSK symbol-streams and the signal separation method that collects none interference between the two BPSK symbols, but gains an extra symbol energy in comparison with the conventional symbol-detection methods.  This is the reason that we can achieve higher reliable data rate beyond the channel capacity in the low SNR regime.   

Throughout the paper, we work on mutual information \eqref{eqR} with AWGN channel model \eqref{eqy} and use $E_{in}=\mathcal{E}(x^2)$ to denote the average input-energy constraint and $\gamma=E_{in}/\sigma_N^2 $ the SNR for the comparison with the channel capacity, where $\mathcal{E}\{\cdot\}$ represents the expecting operator. While, for simplicity, we can use $\tilde{I}_{x_i}(\gamma_i)$ to represent the mutual information $I(X_i;Y_i)$ with $\gamma_i=\mathcal{E}(x_i)^2/\sigma_N^2$ in derivations~\cite{Verdu2007}.     
 
\section{Signal Layering and Splitting}
Consider a single AWGN channel with the input of two independent BPSK symbol-streams of the sufficient long length and the corresponding output. We can specify the above transmission by using the channel model \eqref{eqy} with $\alpha x_1$ and $\beta x_2$ to represent the two symbol-streams, where $\alpha$ and $\beta$ are the two amplitudes with the assumption of $\alpha>\beta>0$, and ${x_1, x_2}\in \{+1, -1\}$ with $+1$ and $-1$ occurring at the equal-probability, respectively. 

The two BPSK symbols are layered at the channel input
\begin{eqnarray}\label{eqx}
x = \alpha x_1+\beta x_2,
\end{eqnarray}
where $x$ is defined as the cocktail BPSK signal, whose average energy is equal to that of input by $E_{in}=\mathcal{E}(x^2)=\alpha^2+\beta^2$.

Taking \eqref{eqx} into \eqref{eqy} and categorizing the equation into two cases; (I) when $x_1=x_2$ and (II) when $x_1=-x_2$, we construct the detection function of $x_1$ by
\begin{eqnarray}\label{eqy1}
y_1 = A_j x_1+n, \qquad j=1 \ \rm{or}\ 2,
\end{eqnarray}
where $y_1$ is used to replace $y$ in the expression for marking the detection of $x_1$, $A_jx_1$ is the BPSK having two amplitudes of equal-probability by $A_1=\alpha+\beta$ and $A_2=\alpha-\beta$ with $j = 1$ and $2$ indicating case I and II, as listed in Table I, respectively.  

We note that the case I and II are independent of each other with same occurring probability.

It is found that the symbol energy we use to the detection of $x_1$ equals to the input-energy by $\frac{1}{2}A_1^2+\frac{1}{2}A_2^2 = E_{in}$. 

\begin{table}[htb]
	\renewcommand{\arraystretch}{1.5}
	\centering
	\small
	\caption{Parallel transmission of cocktail BPSK symbols.}
	\label{Table1}
	\begin{tabular}{c|c|c|c}
		\hline
		\tabincell{c}{Case} & \tabincell{c}{$x_1$} & \tabincell{c}{$x_2$}&\tabincell{c}{$y_1$} \\
		\hline		
		\multirow{2}*{I} & \tabincell{c}{$+1$} & \tabincell{c}{$+1$}& \tabincell{c}{$+(\alpha +\beta) +n$} \\
		\cline{2-4}
		~ & \tabincell{c}{$-1$} &\tabincell{c}{$-1$} & \tabincell{c}{$-(\alpha+\beta)+n$} \\
		\hline		
		\multirow{2}*{II} & \tabincell{c}{$+1$} & \tabincell{c}{$-1$} &\tabincell{c}{$+(\alpha-\beta) +n$} \\
		\cline{2-4}
		~ & \tabincell{c}{$-1$}& \tabincell{c}{$+1$} &\tabincell{c}{$-(\alpha-\beta)+n$} \\
		\hline
		
	\end{tabular}
\end{table}

Upon the detection of $x_1$, we can construct another equation for the detection of $x_2$ by subtracting $x_1$ from \eqref{eqy1}, i.e., 
\begin{eqnarray}\label{eqy2}
y_2 = y_1-\alpha \hat x_1=\beta x_2 + n,   
\end{eqnarray} 
where $y_2$ is used to detect $x_2$, and $\hat x_1$ is the recovered symbol of $x_1$ at the receiver. 

We note that the last equality in right side of \eqref{eqy2} holds as explained in the next subsection, and the symbol energy used to the detection of $x_2$ is $\beta^2$.

Somewhat surprisingly, we can find the total energy used in the above two-step-detection is greater than the input-energy. Actually, an energy gain can be found by  
\begin{eqnarray}\label{eqGe}
G_E = E_{used}-E_{in} = \alpha^2+2\beta^2-E_{in}=\beta^2,   
\end{eqnarray} 
where $E_{used}$ is the total detection-energy and $G_E=\beta^2$ the energy gain which can not be obtained in the most conventional symbol-demodulations, e.g., in the detection of $4$ASK.

\section{Theoretic Proof}
This section proves that the ADR of cocktail BPSK can be higher than the channel capacity at zero SNR limit.  

To work on the calculations of the ADR, we use $z \in \{A_j x_1\}$ to the substitution in \eqref{eqy1} and $ g \in \{\beta x_2\} $ in \eqref{eqy2} and prove  
\begin{eqnarray}\label{eqI}
I(X;Y)= I(Z;Y_1) + I(G;Y_2),
\end{eqnarray}
where $I(X;Y)$ presents the mutual information of \eqref{eqy}, $I(Z;Y_1)$ and $I(G;Y_2)$ \eqref{eqy1} and \eqref{eqy2}, respectively.

It is obvious that the correctness of \eqref{eqI} hinges on the last equality in the right side of \eqref{eqy2}, which will be proved under the condition that the transmission rate owing to $x_1$ is not lager than $I(Z; Y_1)$. 

Proof of \eqref{eqy2}: According to the Shannon information theory, when the data rate owing to the transmission of $x_1$ is not larger than the upper bound $I(Z;Y_1)$, the error free transmission of the source code can be guaranteed. Based on this, the receiver can use the received source code of error-free to recover $x_1$ by performing the channel coding and the symbol mapping in the same manner as that of the transmitter. Thus, $\hat x_1 =x_1 $ holds theoretically and the proof of \eqref{eqI} completes. 

Then, we calculate the ADR owing to the transmission of $x_1$ in \eqref{eqy1}. 

For presenting the derivation more clearly, we re-write \eqref{eqy1} in form of 
\begin{eqnarray}\label{eqyz}
y_z = z + n,
\end{eqnarray}
where $z \in \{z_1,z_2\}$ with $z_1 \in \{A_1 x_1\}$ and $z_2 \in \{A_2 x_1\}$ , and we prove 
\begin{eqnarray}\label{eqyZG}
I(Z;Y_1) = \tilde{I}_{z}(\gamma)=\frac{1}{2}\tilde{I}_{z_1}(\gamma_1) +\frac{1}{2}\tilde{I}_{z_2}(\gamma_2),
\end{eqnarray}
where $\gamma=(\alpha^2+\beta^2)/\sigma_N^2$, $\gamma_1=A_1^2/\sigma_N^2$, and $\gamma_2=A_2^2/\sigma_N^2$, respectively. 

Proof of \eqref{eqyZG}: According to the mutual information theorem [1]: when channel input consists of two independent finite alphabet sets, i.e., $z \in \{z_1,z_2\}$ where $z_1$ and $z_2$ are independent of each other, the mutual information can be calculated by
\begin{eqnarray}\label{eqIzyz}
I(Z;Y_z) = \eta_1 H(Y_{z_1}) + \eta_2 H(Y_{z_2}) + H(N),
\end{eqnarray}
where ${\rm{H}}(Y_{z_i})$ is the entropy corresponding to the signal set represented by $z_i$, for $i=1,2$, ${\rm{H}}(N) = {\log _2} (\sqrt{2 \pi e \sigma_N^2})$ is the entropy of the noise, and $\eta_1$ and $\eta_2$ are the occurring probability of $z_1$ and $z_2$ with $\eta_1+\eta_2=1$, respectively. 

Equation \eqref{eqIzyz} can be re-written in the following derivations 
\begin{eqnarray}\label{eqIzy}
\begin{aligned}
I(Z;Y) &= \eta_1 H(Y_{z_1}) + \eta_1H(N) + \eta_2 H(Y_{z_1}) + \eta_2 H(N) \\
&= \eta_1 {I}(Z_1;Y_{z_1}) + \eta_2 I(Z_2; Y_{z_2})\\
&= \eta_1\tilde{I}_{z_1}(\gamma_1) +\eta_2 \tilde{I}_{z_2}(\gamma_2) ,
\end{aligned}
\end{eqnarray}
where $\gamma_1=(\alpha+\beta)^2/\sigma_N^2$ and $\gamma_2=(\alpha-\beta)^2/\sigma_N^2$.

Note that $z_1$ and $z_2$ present signals of the two cases, i.e., case I and case II, which are independent of each other, we obtain that $\mathcal {E}(z^2)= \eta_1 z_1^2+ \eta_2 z_2^2 $ , thus, $\gamma=\eta_1\gamma_1+\eta_2\gamma_2$. 

Setting $\eta_1=\eta_2=\frac{1}{2}$ in \eqref{eqIzy}, we complete the proof of \eqref{eqyZG}. 

To complete \eqref{eqI} for the calculations, we adds the mutual information inform of \eqref{eqy2} by $ I(G;Y_2) = \tilde{I}_{g}(\gamma_3)$, where $\gamma_3=\beta^2/\sigma_N^2$. 

Finally, \eqref{eqR} can be specified as
\begin{equation}\label{eqR2}
\mathbb{R} = \frac{1}{2}\tilde{I}_{z_1}(\gamma_1)+\frac{1}{2}\tilde {I}_{z_2}(\gamma_2)+ \tilde{I}_g (\gamma_3),
\end{equation}
where the first two terms in right side owning to the transmission of $x_1$ and the last term $x_2$. 

Now we are ready to work on the zero SNR limit for calculating the ADR of cocktail BPSK.

As a preparation, let us recall the following theorem that holds for the mutual information of BPSK/(binary antipode) at the first order approximation at zero SNR~\cite{Shannon1948,Verdu2011}. 

Theorem:

The first order approximation of the mutual information, the result depends only on the SNR and, is independent of the input distribution. The explicit expression can be written as 
\begin{eqnarray}\label{eqIxi}
\tilde{I}_{x_i}(\gamma_i) = C(\gamma_i) + O(\gamma_i), \ \  \rm{for}\ \ \gamma_i \to 0 ,
\end{eqnarray}
where $C$ is the channel capacity, $I_{x_i}$ is a mutual information with input $x_i$ and $\gamma_i=\mathcal{E}(x_i^2)/\sigma_N^2$ is the corresponding SNR.

By using Taylor's expansion to the function of the channel capacity, \eqref{eqIxi} can be written as
\begin{eqnarray}\label{eqIxd}
\tilde{I}_{x_i}(\gamma_i) = C'(0)\gamma_i + O(\gamma_i), \ \  \rm{for}\ \ \gamma_i \to 0 ,
\end{eqnarray}
where $C'(0)$ represents the first derivative of the channel capacity.

We define the low-SNR region by $\gamma_1=(\alpha+\beta)^2/\sigma_N^2<<1$ that is sufficient condition for $\gamma=E_{in}/\sigma_N^2<<1$, and prove the following two equalities in the defined low-SNR region. 

Equality 1
\begin{eqnarray}\label{eqIzy12}
I(Z;Y_1)=\tilde{I}_{x_1}(\gamma)= C(\gamma) ,
\end{eqnarray} 
where $I(Z;Y_1)$ is the mutual information of \eqref{eqy1}, i.e., owing to the transmission of $x_1$.

Equality 2 
\begin{eqnarray}\label{eqIC}
\mathbb{R}-C(\gamma) =\frac{\beta^2}{\sigma_N^2}\log_2e > 0,     
\end{eqnarray}  
where $\mathbb{R}=I(X;Y)$ in \eqref{eqR} is the ADR of cocktail BPSK.  

Proof of equality 1:

Employing \eqref{eqIxi} to \eqref{eqIzy} with the consideration of \eqref{eqIxd} proves \eqref{eqIzy12} by 
\begin{eqnarray}\label{eqIzy13}
\begin{aligned}
I(Z;Y_1) &= \frac{1}{2}C'(0)(\gamma_1+\gamma_2) \\
&= C'(0)\gamma \\
&= C(\gamma).  
\end{aligned}  
\end{eqnarray}
The Proof is completed.

Proof of equality 2: 

Considering \eqref{eqI} and \eqref{eqIzy12}, we can prove \eqref{eqIC} by
\begin{eqnarray}\label{eqIC2}
\begin{aligned}
\mathbb{R} - C(\gamma) &= I(X;Y)-I(Z;Y_1)\\
&= I(G;Y_2)\\
&= \tilde{I}_{x_3}(\gamma_3)\\
&= C'(0)\gamma_3\\
&= \log_2e(\beta^2/\sigma_N^2)>0.
\end{aligned}     
\end{eqnarray}
We complete the proof of the higher ADR of cocktail BPSK beyond the channel capacity. 

It is interesting to note that the spectral efficiency gain in \eqref{eqIC2} is found proportional to $\beta^2$ that is the energy gain in \eqref{eqGe}.

\begin{figure}[ht]
	\centering
	\includegraphics[width=0.5\textwidth]{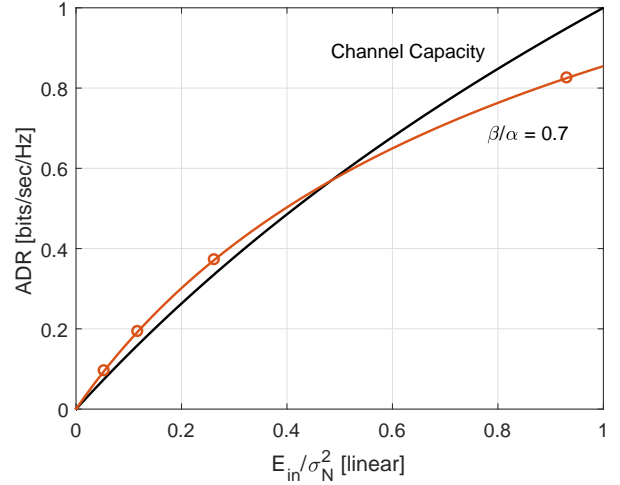}
	\caption{The ADR of cocktail BPSK.}
	\label{fig1}
\end{figure}

\section{Numerical Confirmation}

To confirm the higher ADR of cocktail BPSK beyond the channel capacity, we plot numerical results of \eqref{eqI} with respect to $\gamma=E_{in}/\sigma_N^2$ in the linear measurement at the horizontal axis as shown in Fig.1, where we use $\beta/\alpha = 0.7$ as an example. One can find the gain in the region of SNR = 0 to 0.47.

\end{document}